\documentclass[aps,prd,showpacs,twocolumn,groupedaddress]{revtex4}
\usepackage{graphicx}

\begin{document}

\title{Relevant energy scale of color confinement from lattice QCD}

\author{Arata~Yamamoto}
\affiliation{Department of Physics, Faculty of Science, Kyoto University, Kitashirakawa, Sakyo, Kyoto 606-8502, Japan}

\author{Hideo~Suganuma}
\affiliation{Department of Physics, Faculty of Science, Kyoto University, Kitashirakawa, Sakyo, Kyoto 606-8502, Japan}

\date{\today}

\begin{abstract}
We propose a new lattice framework to extract the relevant gluonic energy scale of QCD phenomena which is based on a ``cut" on link variables in momentum space.
This framework is expected to be broadly applicable to all lattice QCD calculations.
Using this framework, we quantitatively determine the relevant energy scale of color confinement,
through the analyses of the quark-antiquark potential and meson masses.
The relevant energy scale of color confinement is found to be below 1.5 GeV in the Landau gauge.
In fact, the string tension is almost unchanged even after cutting off the high-momentum gluon component above 1.5 GeV.
When the relevant low-energy region is cut, the quark-antiquark potential is approximately reduced to a Coulomb-like potential, and each meson becomes a quasi-free quark pair.
As an analytical model calculation, we also investigate the dependence of the Richardson potential on the cut, and find the consistent behavior with the lattice result.
\end{abstract}


\maketitle

\section{Introduction}
In nature, there exist a large number of physical phenomena, and their scales range from the smallest scale (the Planck scale) to the largest scale (universe size).
These phenomena are described in various ways depending on their scales.
For example, small-scale phenomena are described by elementary particle physics, nuclear physics, and so on, while large-scale phenomena by cosmology, astrophysics, and so on.
The scale characterizes each physical phenomenon and creates their hierarchy.
In modern physics, the scale is one of the most fundamental and important concepts.

In quantum chromodynamics (QCD) and some other field theories, there exists a characteristic phenomenon that describes the nontrivial appearance of a scale.
In classical QCD Lagrangian, only dimensional quantities are quark masses, and therefore QCD is scale invariant in the chiral limit at the classical level.
However, after the quantization, scale invariance is violated and an energy scale appears in a nontrivial manner.
This is called as dimensional transmutation, through which many dimensional quantities are created in real QCD, such as ``mass gap", string tension, various condensates, and mass and size of hadrons.
Dimensional transmutation is an important phenomenon in modern field theory \cite{Co73,Gr74}.
One of the most well-known energy scales in QCD is the QCD scale parameter $\Lambda_{\rm QCD}$, which is defined through the running coupling constant $\alpha_s(Q^2)$ in perturbative QCD.
It is the energy scale where the naive perturbative calculation does not work at all.

The QCD running coupling constant depends on the energy scale as
\begin{eqnarray}
\alpha_s(Q^2) =\frac{1}{4\pi\beta_0 \ln (Q^2/\Lambda_{\rm QCD}^2)},
\end{eqnarray}
where $\beta_0=(11N_c-2N_f)/48\pi^2$, at one-loop level.
Therefore, the behavior of QCD phenomena at high energy is entirely different from that at low energy.
At high energy or short distance, perturbative QCD is valid due to the asymptotic freedom \cite{Gr73,Po73}.
In contrast, at low energy or long distance, the value of the QCD running coupling constant is large and nonperturbative effects are important.
There exist many characteristic phenomena in nonperturbative region, such as quark confinement and chiral symmetry breaking \cite{Gr03,Ha99,Hi91,Co82,Ch08}.
It is found that the appearance of the energy scale enriches the QCD phenomena.

One important question is {\it what energy scale is relevant for each QCD phenomenon}.
One may naively believe that the typical energy scale of nonperturbative QCD is $\Lambda_{\rm QCD}$.
However, $\Lambda_{\rm QCD}$ is just one scale that is defined through the running coupling constant.
We do not know which energy components of gluons and quarks mainly contribute to a QCD phenomenon.
Also, it is not trivial whether such relevant energy components are the same or different between different nonperturbative phenomena, for example, quark confinement and chiral symmetry breaking.

In order to determine the relevant energy scale of QCD phenomena, we employ lattice QCD, which is the nonperturbative and first-principle calculation of QCD \cite{Cr81,Ro92}.
In this paper, we propose a new framework in lattice QCD to determine the relevant gluonic energy scale systematically, quantitatively, and nonperturbatively \cite{YaL}.
We study the relevant gluonic energy scale of color confinement by this framework.

The outline of the paper is as follows.
In Sec.~II, we formulate a lattice framework to determine the relevant energy scale for QCD phenomena.
This framework is fairly general and readily applicable to actual lattice calculations.
To investigate the relevant energy scale of color confinement, we calculate two fundamental examples of quenched lattice calculations; the quark-antiquark potential in Sec.~III and meson masses in Sec.~IV.
In Sec.~V, we perform an analytical study for the relevant energy scale of color confinement.
We investigate the dependence of the confinement in the Richardson potential on the value of the infrared cut, and compare it with the lattice results presented in Sec.~III.
Finally, Sec.~VI is devoted to summary and discussion.

\section{Lattice Formulation}
The main concept of our framework is simple and general.
It is to calculate a physical quantity, after artificially cutting a certain region of momentum space.
From the relationship between the cut region and the resulting quantity, we can determine the relevant energy scale of the quantity.
This concept would be applicable to many theories in physics \cite{Ii05}.
Here, we introduce this concept to lattice QCD.
The fundamental degree of freedom in lattice QCD is a link variable, which represents a gluon field on lattice.
We consider a ``cut" on link variables in momentum space, and investigate its effect on the resulting quantity.

The procedure for each gauge configuration is formulated as the following five steps.

Step 1. We generate a gauge configuration by Monte Carlo simulation of lattice QCD under space-time periodic boundary conditions, and obtain a finite number of coordinate-space link variables
\begin{equation}
U_{\mu}(x)=e^{iagA_\mu (x)}\in {\rm SU(3)},
\end{equation}
where $a$ is the coordinate-space lattice spacing.
Since the procedure is not gauge invariant, we fix the link variables with some gauge.
It is desirable to choose a physically meaningful gauge on the lattice.
In this paper, we mainly use the Landau gauge.
The Landau gauge is well-known and frequently used also in continuum theory, and it gives a transparent connection between the link variable and the gauge field.

Step 2. By carrying out discrete Fourier transformation in the four-dimensional Euclid space, we define the momentum-space link variable ${\tilde U}_{\mu}(p)$ as
\begin{eqnarray}
{\tilde U}_{\mu}(p)=\frac{1}{N_{\rm site}}\sum_x U_{\mu}(x)\exp(i {\textstyle \sum_\nu} p_\nu x_\nu),
\end{eqnarray}
where $N_{\rm site}$ is the total number of lattice sites.
The momentum space represents a lattice with $N_{\rm site}$ lattice sites, and boundary conditions are periodic.
The first Brillouin zone of the momentum space is a four-dimensional hypercube with each side of $(-\pi/a, \pi/a ]$.
The momentum-space lattice spacing is given by
\begin{eqnarray}
a_p = \frac{2\pi}{La},
\end{eqnarray}
where $L$ is the number of lattice sites in each direction.
The momentum-space lattice spacing corresponds to the minimum unit of momentum, and it has mass dimension.

\begin{figure}[t]
\begin{center}
\includegraphics[scale=0.5]{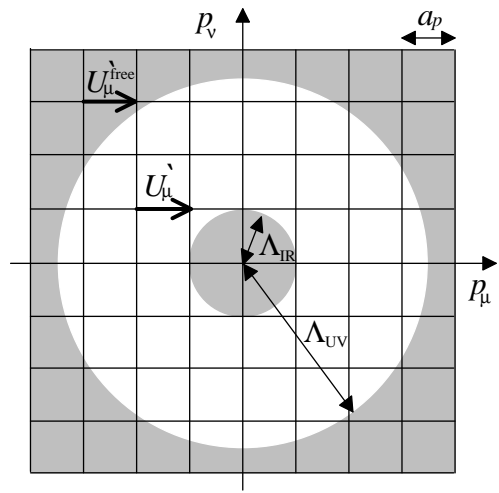}
\caption{\label{Fig1}
Two-dimensional schematic figure of the UV cut $\Lambda_{\rm UV}$ and the IR cut $\Lambda_{\rm IR}$ on momentum-space lattice.
The cut (shaded) region is described by Eq.~(\ref{eq5}).
The momentum-space link variable ${\tilde U}_{\mu}(p)$ is replaced with the free variable ${\tilde U}^{\rm free}_{\mu}(p)$ in the cut regions.
$a_p$ is the momentum-space lattice spacing.
}
\end{center}
\end{figure}

Step 3. We introduce a ``cut" on ${\tilde U}_{\mu}(p)$ in a certain region of momentum space.
Outside the cut, ${\tilde U}_{\mu}(p)$ is replaced by the free-field link variable
\begin{equation}
{\tilde U}^{\rm free}_{\mu}(p)=\frac{1}{N_{\rm site}}\sum_x 1 \exp(i {\textstyle \sum_\nu} p_\nu x_\nu)=\delta_{p0}.
\end{equation}
Then, the momentum-space link variable with the cut is defined as 
\begin{equation}
\label{eq2}
{\tilde U}_{\mu}^{\Lambda}(p)= \Bigg\{
\begin{array}{cc}
{\tilde U}_{\mu}(p) & ({\rm inside \ cut})\\
{\tilde U}^{\rm free}_{\mu}(p)=\delta_{p0} & ({\rm outside \ cut}).
\end{array}
\end{equation}
The concrete form of the cut can be taken arbitrarily.
The most natural choice to estimate the energy scale is the cut by a four-momentum length $\sqrt{p^2}=\sqrt{\sum_\mu p_\mu p_\mu}$, which corresponds to a simple momentum cut in continuum theory.
For example, an ultraviolet (UV) cut or an infrared (IR) cut by a four-momentum length is
\begin{equation}
\label{eq5}
\sqrt{p^2} > \Lambda_{\rm UV} \ {\rm or} \ \sqrt{p^2} < \Lambda_{\rm IR}.
\end{equation}
These cuts are schematically depicted in Fig.~\ref{Fig1}.
In this study, we mainly use these UV and IR cuts for numerical calculations.
Another possible choice is, for example, ${\rm max}(p_1,p_2,p_3,p_4)$ instead of $\sqrt{p^2}$, which respects the lattice structure.
Of course, we can cut not only the UV or IR region, but also intermediate-momentum region, anisotropic region, and so on.

Step 4. To return to coordinate space, we carry out the inverse Fourier transformation as
\begin{eqnarray}
U'_{\mu}(x)=\sum_p {\tilde U}_{\mu}^{\Lambda}(p)\exp(-i {\textstyle \sum_\nu} p_\nu x_\nu).
\end{eqnarray}
Since this $U'_{\mu}(x)$ is not an SU(3) matrix, we project it onto an SU(3) element $U^{\Lambda}_{\mu}(x)$ by maximizing
\begin{eqnarray}
{\rm ReTr}[U^{\Lambda}_{\mu}(x)^{\dagger}U'_{\mu}(x)].
\end{eqnarray}
Such a projection is often used in lattice QCD algorithms.
By this projection, we obtain the coordinate-space link variable $U^{\Lambda}_{\mu}(x)$ with the cut, which is an SU(3) matrix and has the maximal overlap to $U'_{\mu}(x)$.

Step 5. Using these link variables $U^{\Lambda}_{\mu}(x)$, we compute the expectation value of physical quantities in the exact same way as original lattice QCD.

By repeating Steps 3-5 with different values of the cut, we can quantitatively determine the relevant energy scale of a physical quantity.
Since we only have to replace $U^{\Lambda}_{\mu}(x)$ instead of $U_{\mu}(x)$, this framework can be applied to all lattice calculations.
In addition, since the fast Fourier transformation is numerically easy task compared to gauge configuration generation or solver calculation, the computing time required for each cut is almost the same as that required for original lattice calculations.
We expect that this framework can be broadly used to the analysis for the relevant energy scale of QCD phenomena.

As for the gauge fixing, we mainly use the Landau gauge.
In lattice QCD with the Euclidean metric, the Landau gauge is defined by the condition that globally maximizes the quantity
\begin{eqnarray}
F[U] \equiv \sum_{x}\sum_{\mu} {\rm ReTr} U_\mu (x),
\end{eqnarray}
by the SU(3) gauge transformation.
In terms of the gauge field $A_\mu (x)$, this condition is equivalent to minimizing 
\begin{eqnarray}
\int d^4x {\rm Tr}\{ A_\mu (x)^2 \}
\end{eqnarray}
in Euclidean QCD, and it is a sufficient condition for the local condition $\partial_\mu A_\mu (x)=0$.
Then, the gauge fluctuation is maximally suppressed, so that the link variable can be expanded as
\begin{eqnarray}
U_{\mu}(x)=1+igaA_\mu (x)+\cdots
\end{eqnarray}
in a well-defined manner, and there exists a transparent connection between the link variable and the gauge field.
If one is interested in the direct information about the gauge field, it is possible to construct the similar framework with $gA_\mu (x)$, which is renormalization-group invariant, instead of $U_{\mu}(x)$.
For example, in the Landau gauge, $gA_\mu (x)$ is approximately extracted as
\begin{eqnarray}
gA_\mu (x) &=& g\bar{A}_\mu (x) - \frac{g}{N_c}{\rm Tr} \bar{A}_\mu (x),\\
g\bar{A}_\mu (x)&\equiv& \frac{1}{2ia}(U_{\mu}(x) - U_{\mu}^{\dagger}(x)).
\end{eqnarray}
The trace part is subtracted for the traceless property of the gluon field.

One may feel that our framework is similar to renormalization.
This framework is {\it not} equivalent to renormalization transformation.
In the case of renormalization transformation, the gauge field outside cutoff is integrated out, however, in the case of our framework, the gauge field outside the cut is simply removed.
Such a simple removal of the gauge-field components breaks gauge invariance.
Since the gauge transformation property is nonlocal and nontrivial in momentum space, gauge invariance is lost when a certain momentum region is cut. 
If we do not fix the gauge, the resulting expectation value is zero, even in the case of a gauge-invariant operator.
Then, gauge fixing is needed in the framework.
Note, however, that the procedure itself is not restricted to a specific gauge, and we can use other gauges instead of the Landau gauge.
By comparing the results obtained in several different gauges, we can check the gauge dependence of the relevant energy scale.

In the case of a general gauge, since gauge fixing and momentum cut do not necessarily commute, it is nontrivial that these two conditions are satisfied simultaneously.
In other words, $U^{\Lambda}_{\mu}(x)$ can deviate from the gauge-fixing condition which is originally imposed on $U_{\mu}(x)$.
However, in the Landau gauge, we have numerically checked that $U^{\Lambda}_{\mu}(x)$ almost satisfies the Landau gauge fixing condition, i.e., $F[U^{\Lambda}]$ is maximized, even after the back projection to SU(3).

\section{Interquark potential}
We apply our framework to the analysis of the interquark potential in terms of gluonic energy components.
The interquark potential is one of the most fundamental quantities in QCD, and many types of interquark potentials have been studied by lattice QCD \cite{Cr7980,Ta0102,Ok05,Ya08}.
Among them, we here calculate the most basic potential, i.e., the quark-antiquark ($Q\bar Q$) potential.
The $Q\bar Q$ potential is extracted from the expectation value of the Wilson loop, which is a gauge-invariant path-ordered product of link variables along a loop.

\begin{figure}[b]
\begin{center}
\includegraphics[scale=1.15]{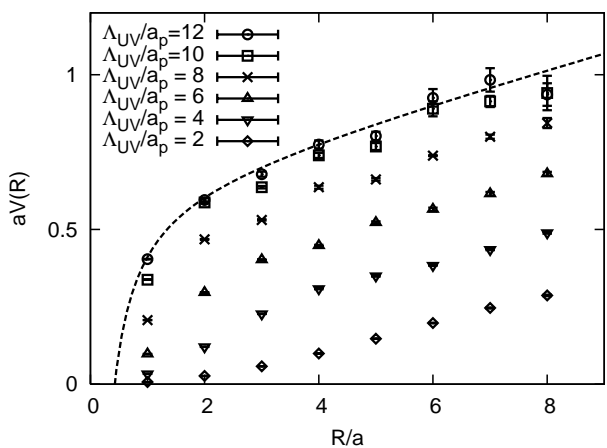}
\caption{\label{Fig2}
The $Q\bar Q$ potential with the UV cut $\Lambda_{\rm UV}$ plotted against the interquark distance $R$.
The lattice QCD calculation is performed on $16^4$ lattice with $\beta =6.0$.
The unit is scaled with the lattice spacing, $a\simeq 0.10$ fm or $a_p \simeq 0.77$ GeV.
The broken line is the original $Q\bar Q$ potential in lattice QCD.
}
\end{center}
\end{figure}

\begin{table*}[t]
\newcommand{\m}{\hphantom{$-$}}
\newcommand{\cc}[1]{\multicolumn{1}{c}{#1}}
\renewcommand{\tabcolsep}{1pc} 
\renewcommand{\arraystretch}{1} 
\caption{\label{tab1}
Asymptotic string tension $\sigma_{\rm asym}$ of the interquark potential with the IR cut $\Lambda_{\rm IR}$.
The results obtained under four different conditions are listed; the main result (with the four-dimensional hyperspherical cut in the Landau gauge) obtained in Sec.~III-A, the result in the Coulomb gauge, the result with the hypercubic cut, and the result with three-dimensional formalism obtained in Sec.~III-C.
The lattice QCD calculations are performed on $16^4$ lattice with $\beta =6.0$, where the lattice spacings are $a\simeq 0.10$ fm and $a_p \simeq 0.77$ GeV.
In original lattice QCD, the string tension $\sigma a^2$ is about 0.051.
}
\begin{center}
\begin{tabular}{ccccc}
\hline\hline
$\Lambda_{\rm IR}/a_p$ &$\sigma_{\rm asym}a^2$ & $\sigma_{\rm asym}a^2$ (Coulomb)& $\sigma_{\rm asym}a^2$ (hypercube)&$\sigma_{\rm asym}a^2$ (3-dim.)\\
\hline
1.0 & 0.0469(58) & 0.0289(58) & 0.0469(58)& 0.0433(51)\\
1.1 & 0.0311(49) & 0.0190(59) & -& 0.0198(43)\\
1.5 & -0.0019(20)& 0.0024(25) & -& -0.0034(9)\\
2.0 & -0.0132(6) & -0.0041(10)& -0.0142(8)& -0.0092(5)\\
3.0 & 0.0003(12) & 0.0058(10) & 0.0065(20)& 0.0024(13)\\
\hline\hline
\end{tabular}
\end{center}
\end{table*}

\subsection{$Q\bar Q$ potential with UV/IR cut}
The $Q\bar Q$ potential is expressed as a sum of one-gluon-exchange Coulomb potential and linear confinement potential as
\begin{eqnarray}
V(R)=\sigma R -\frac{A}{R} +C,
\label{VQQ}
\end{eqnarray}
where $R$ is the distance between quark and antiquark.
The physical value of the string tension $\sigma$ is approximately 0.89 GeV/fm, and the Coulomb coefficient $A$ is approximately 0.26.
The constant $C$ is physically irrelevant, and its value depends on regularization.
The $Q\bar Q$ potential includes both perturbative and nonperturbative ingredients.
In short range, it is dominated by the perturbative one-gluon-exchange Coulomb potential.
In long range, it is dominated by the linear confinement potential, which is purely a nonperturbative phenomenon.
We determine the relevant gluonic energy scales of the Coulomb and confinement potentials.

The numerical simulation in this subsection is performed with the isotropic plaquette gauge action with $\beta=6.0$.
The lattice size is $16^4$ and periodic boundary condition is imposed, and then the corresponding momentum-space lattice is also a $16^4$ isotropic lattice.
The gauge configuration number is 50, however, for statistical improvement, we average all the parallel-translated Wilson loops in one configuration and apply the APE smearing method \cite{Al87}.
The coordinate-space lattice spacing $a$ is about 0.10 fm, and the lattice volume is $(1.6 \ {\rm fm})^4$.
The momentum-space lattice spacing $a_p$ is about 0.77 GeV, and the momentum-space lattice volume is $(12 \ {\rm GeV})^4$.
In this paper, a dimensional quantity in coordinate space is scaled with $a$ by the standard lattice convention.
Similarly, a dimensional quantity in momentum space, such as the cut $\Lambda_{\rm UV}$ or $\Lambda_{\rm IR}$, is scaled with $a_p$.

First, we introduce the UV cut $\Lambda_{\rm UV}$ to the $Q\bar Q$ potential.
The resulting potential is shown in Fig.~\ref{Fig2}.
Since the maximum value of the four-momentum length is $\sqrt{p^2}/a_p=16$ in our lattice, the standard lattice result (broken line) corresponds to the case of $\Lambda_{\rm UV}/a_p=16$.
When the value of the UV cut becomes smaller, the short-range Coulomb potential gradually decreases.
The constant term $C$ also decreases by the UV cut.
This is because the constant term is mainly given by the lattice regularization for the UV singularity of the Coulomb potential.
On the other hand, the long-range linear potential is almost unaffected by the UV cut.
When we fit the $Q\bar Q$ potential by Eq.~(\ref{VQQ}), the string tension $\sigma$ is insensitive to the value of the UV cut.
At $\Lambda_{\rm UV}/a_p=2$, the $Q\bar Q$ potential becomes only the linear potential, and its fitting result is $\sigma a^2=0.0459(15)$, $A=-0.056(5)$, and $Ca=-0.095(7)$.
The string tension differs slightly from its original value $\sigma a^2\simeq 0.051$, and the Coulomb coefficient and the constant term are nearly zero.

\begin{figure}[b]
\begin{center}
\includegraphics[scale=1.15]{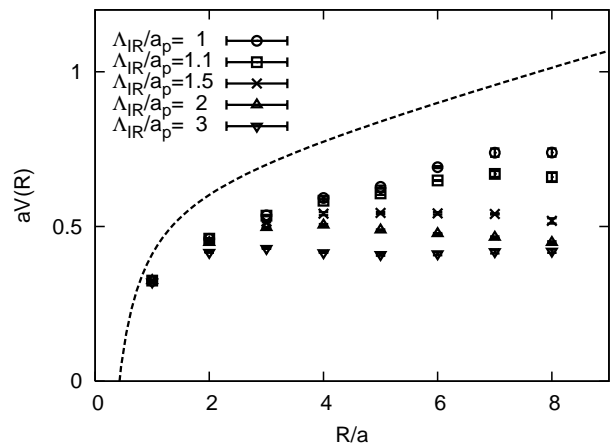}
\caption{\label{Fig3}
The $Q\bar Q$ potential $V(R)$ with the IR cut $\Lambda_{\rm IR}$.
The notation is the same as Fig.~\ref{Fig2}.
}
\end{center}
\end{figure}

Next, we show the $Q\bar Q$ potential with the IR cut $\Lambda_{\rm IR}$ in Fig.~\ref{Fig3}, and list the value of the asymptotic string tension $\sigma_{\rm asym}$ in the second column of Table \ref{tab1}.
The asymptotic string tension $\sigma_{\rm asym}$ is estimated by fitting the $Q\bar Q$ potential in $3< R/a < 9$ with a linear function $\sigma_{\rm asym}R + {\rm const}$.
In contrast to the UV case, the long-range linear potential is affected by the IR cut.
At $\Lambda_{\rm IR}/a_p=1$ and 1.1, the string tension decreases slightly from its original value $\sigma\simeq 0.89$ GeV/fm.
In $\Lambda_{\rm IR}/a_p\ge 1.5$, the long-range linear potential disappears, and the $Q\bar Q$ potential becomes the perturbative Coulomb potential.
Because $a_p\simeq 0.77$ GeV, the physical energy scale for the vanishing of color confinement is about 1 GeV.
From Fig.~\ref{Fig3} and Table \ref{tab1}, we observe that the asymptotic string tension is a small negative value at $\Lambda_{\rm IR}/a_p=1.5$ and 2.
This would suggest that, although the confinement potential asymptotically disappears, it slightly survives only in the intermediate range.
At $\Lambda_{\rm IR}/a_p=3$, the $Q\bar Q$ potential becomes completely flat in $R/a>3$.

From these results, we conclude as the following.
The perturbative and nonperturbative parts of the $Q\bar Q$ potential are decoupled in the momentum space of the gluon.
The relevant energy scale of confinement is below about 1 GeV in the Landau gauge.
If the gluon is restricted to this energy region, the $Q\bar Q$ potential becomes only a linear potential.
In contrast, if this gluon is cut, a Coulomb-like potential is obtained.

\subsection{More quantitative estimate for the energy scale}

\begin{table}[b]
\newcommand{\m}{\hphantom{$-$}}
\newcommand{\cc}[1]{\multicolumn{1}{c}{#1}}
\renewcommand{\tabcolsep}{1pc} 
\renewcommand{\arraystretch}{1} 
\caption{\label{tab2}
Lattice spacings with different lattice couplings $\beta=2N_c/g^2$.
The coordinate-space lattice spacing $a$ and momentum-space lattice spacing $a_p$ are listed.
}
\begin{center}
\begin{tabular}{cccc}
\hline\hline
$\beta$ & lattice size & $a$ [fm] & $a_p$ [GeV]\\
\hline
5.7 & $16^4$ & 0.19 & 0.41\\
5.8 & $16^4$ & 0.14 & 0.55\\
6.0 & $16^4$ & 0.10 & 0.77\\
\hline\hline
\end{tabular}
\end{center}
\end{table}

For a more quantitative argument, we need higher accuracy in momentum space.
Since the minimum momentum on lattice is the momentum-space lattice spacing $a_p$, the four-momentum length $\sqrt{p^2}$ is restricted to discrete values as
\begin{eqnarray}
\sqrt{p^2}=\sqrt{\sum_{\mu}p_\mu p_\mu}=0, a_p, \sqrt{2}a_p, \sqrt{3}a_p, \cdots.
\end{eqnarray}
Then, we can only take discrete variation on the value of the cut.
For example, the lattice calculations in the range $0<\Lambda_{\rm IR}\le a_p$ yield the same result.
This is a kind of discretization error in momentum space.
In order to achieve a finer resolution in momentum, we must calculate with a smaller momentum-space lattice spacing, i.e., a larger coordinate-space lattice volume.
In addition, it is desirable to calculate with several different momentum-space lattice spacings.

\begin{figure}[t]
\begin{center}
\includegraphics[scale=1.15]{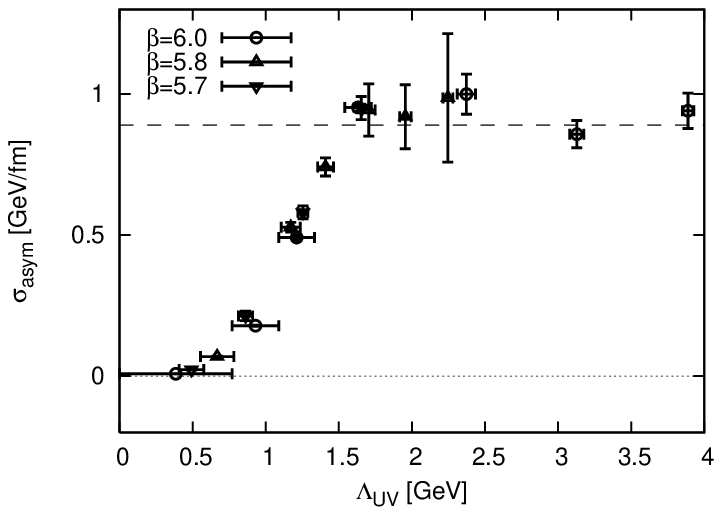}
\caption{\label{Fig4}
The $\Lambda_{\rm UV}$-dependence of the asymptotic string tension $\sigma_{\rm asym}$ in physical unit.
The vertical error bar is the standard statistical error, and the horizontal error bar is the range that yields the same result due to the discrete momentum.
The original value of the string tension is $\sigma \simeq 0.89$ GeV/fm (broken line).
}
\includegraphics[scale=1.15]{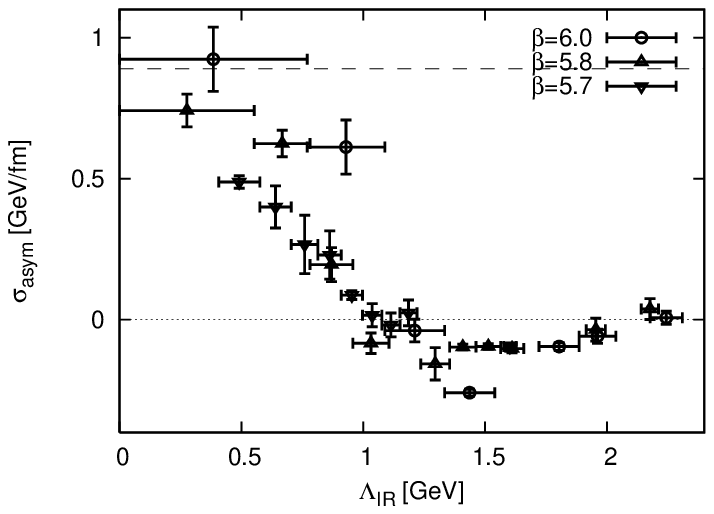}
\caption{\label{Fig5}
The $\Lambda_{\rm IR}$-dependence of the asymptotic string tension $\sigma_{\rm asym}$ in physical unit.
The notation is the same as that used in Fig.~\ref{Fig4}.
}
\end{center}
\end{figure}

For more quantitative estimate on the relevant energy scale of color confinement, we calculate the $Q\bar Q$ potential on $16^4$ lattice with $\beta=5.7$, 5.8, and 6.0.
As listed in Table \ref{tab2}, the corresponding momentum-space lattice spacings are $a_p\simeq 0.41$ GeV, 0.55 GeV, and 0.77 GeV, respectively.
The coordinate-space lattice spacings are determined so as to reproduce the string tension $\sigma$ to be 0.89 GeV/fm.
Other conditions are the same as before.

In Fig.~\ref{Fig4}, we show the $\Lambda_{\rm UV}$-dependence of the asymptotic string tension $\sigma_{\rm asym}$.
The asymptotic string tension $\sigma_{\rm asym}$ is estimated by fitting the $Q\bar Q$ potential with a linear function $\sigma_{\rm asym}R + {\rm const.}$ in $0.3 \ {\rm fm}< R < 0.9$ fm.
In this figure, while the error bar in the vertical direction represents the standard statistical error, the error bar in the horizontal direction represents not the statistical error but the range which yields the same result due to the discrete momentum.
The asymptotic string tension is almost unaffected in $\Lambda_{\rm UV}> 1.5$ GeV, and its value is unchanged from its original value $\sigma \simeq 0.89$ GeV/fm.
In $\Lambda_{\rm UV}< 1.5$ GeV, the value of the asymptotic string tension significantly decreases.

We analyze the IR case in the same way and show the result in Fig.~\ref{Fig5}.
All the results consistently suggest that the energy scale of the vanishing of the asymptotic string tension is around $\Lambda_{\rm IR}=1.2$ GeV.
Above this energy scale, the asymptotic string tension is almost zero and the confinement potential asymptotically disappears .

From these quantitative analyses, we conclude that the relevant energy scale of color confinement is below 1.5 GeV, i.e., color confinement originates from the low-energy gluon components below 1.5 GeV.

\subsection{Consistency check with other conditions}

\begin{figure}[t]
\begin{center}
\includegraphics[scale=0.5]{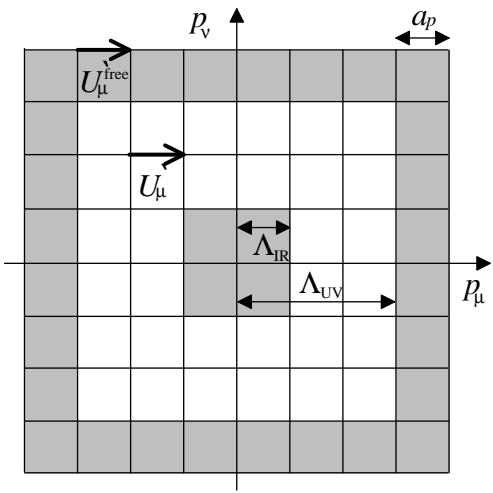}
\caption{\label{Fig6}
Another example of a cut on momentum-space lattice.
The cut (shaded) region is described by Eq.~(\ref{eq6}).
Other notations are the same as those used in Fig.~\ref{Fig1}.
}
\end{center}
\end{figure}

We have calculated with other conditions for the check of consistency: (i) a different gauge in Step 1, (ii) a different cut in Step 3, and (iii) the three-dimensional Fourier transformation.
The results obtained under these three conditions are consistent with the previous results.
Here, we briefly summarize these calculations and results.

(i) Since our framework is not gauge invariant, it is important to check that the resulting energy scale does not depend drastically on the gauge choice.
We calculate with the Coulomb gauge for the gauge choice in the Step 1, instead of the Landau gauge.
The gauge-fixing condition for the Coulomb gauge is to maximize the quantity
\begin{eqnarray}
\sum_{x} \sum_{j=1}^{3} {\rm ReTr} U_j (x).
\end{eqnarray}
The values of $\sigma_{\rm asym}$ with the Coulomb gauge are listed in the third column of Table \ref{tab1}.
The value itself depends on the gauge choice, however, the energy scale for the vanishing of confinement is the same, that is, about 1 GeV.

(ii) Our framework has an ambiguity of the cut form.
For example, since the rotational invariance is broken on lattice, it is worth introducing a cut which respects the lattice structure of momentum space.
Then we calculate with the cut by ${\rm max}(p_1,p_2,p_3,p_4)$ instead of $\sqrt{p^2}$, i.e.,
\begin{equation}
\label{eq6}
{\rm max}(p_1,p_2,p_3,p_4) > \Lambda_{\rm UV} \ {\rm or} \ {\rm max}(p_1,p_2,p_3,p_4) < \Lambda_{\rm IR},
\end{equation}
instead of Eq.~(\ref{eq2}).
This cut forms a four-dimensional hypercube, as shown in Fig.~\ref{Fig6}.
The values of $\sigma_{\rm asym}$ with this cut are listed in the fourth column of Table \ref{tab1}.
The result estimated by ${\rm max}(p_1,p_2,p_3,p_4)$ is consistent with that estimated by $\sqrt{p^2}$.

(iii) Our framework can be easily extended to the spatial three-dimensional formalism.
The spatial three-dimensional formalism is related to the Richardson potential calculation, which is shown in Sec.~V.
We only have to perform the three-dimensional Fourier transformation and consider a cut by the three-momentum length $|\vec{p}|$.
The result of the three-dimensional formalism is listed in the fifth column of Table \ref{tab1}.
In the Landau gauge, the relevant three-dimensional momentum scale
of confinement is found to be below about 1 GeV, which is almost
the same as the relevant four-dimensional energy scale.

\section{Meson masses}
In this section, we apply our framework to the analysis of meson masses in quenched lattice QCD.
We calculate the mass of a pion (pseudo-scaler meson) and a $\rho$-meson (vector meson) using two different fermion actions, i.e., the clover fermion action and the staggered fermion action.

\subsection{Clover fermion}

The clover fermion action is an $O(a)$-improved Wilson fermion action, and the quark mass is described in terms of the mean-field-improved hopping parameter $\kappa$ \cite{Ya08,Wi75,Le93,El97}.
For the details of the improvement, see Ref.~\cite{Ya08}.
The Wilson fermion is the simplest fermion on lattice, but it explicitly breaks chiral symmetry for avoiding the doubling problem.

The numerical simulation is performed on $16^4$ lattice with $\beta=6.0$.
The lattice spacings are the same as before, i.e., $a\simeq$ 0.10 fm and $a_p\simeq$ 0.77 GeV.
Three different values of hopping parameters are taken as $\kappa=0.1200$, 0.1300, and 0.1340, and the corresponding pion masses are about 2.9 GeV, 1.8 GeV, and 1.3 GeV, respectively.
The configuration number is 100 here.

\begin{figure}[t]
\begin{center}
\includegraphics[scale=1.15]{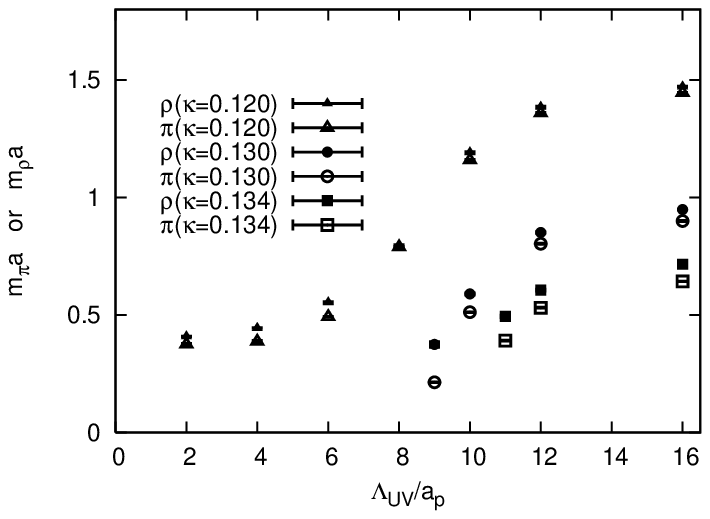}
\caption{\label{Fig7}
The $\Lambda_{\rm UV}$-dependence of pion mass $m_\pi$ and $\rho$-meson mass $m_\rho$.
The quark propagator is calculated by the clover fermion action with the hopping parameter $\kappa$.
The data at $\Lambda_{\rm UV}/a_p=16$ is the standard lattice result.
The unit is scaled with the lattice spacing, $a\simeq 0.10$ fm or $a_p \simeq 0.77$ GeV.
}
\includegraphics[scale=1.15]{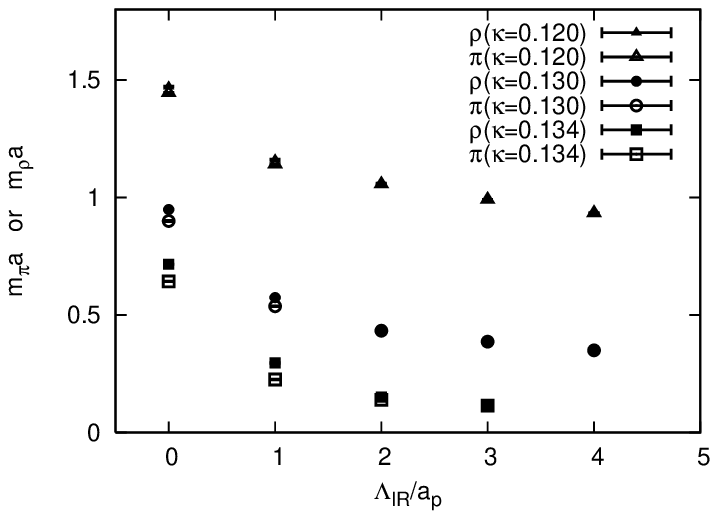}
\caption{\label{Fig8}
The $\Lambda_{\rm IR}$-dependence of meson masses with the clover fermion.
The data at $\Lambda_{\rm IR}/a_p=0$ is the standard lattice result.
The notation is the same as Fig.~\ref{Fig7}.
}
\end{center}
\end{figure}

Figure \ref{Fig7} shows the pion mass $m_\pi$ and the $\rho$-meson mass $m_\rho$ with the UV cut.
In this figure, the data at the far right side, i.e., $\Lambda_{\rm UV}/a_p=16$, is the standard lattice result.
As the value of the UV cut becomes smaller, both the pion mass and the $\rho$-meson mass gradually decreases.
This decrease corresponds to the mass generated by the UV gluon which dresses the quarks in the mesons.

Figure \ref{Fig8} shows $m_\pi$ and $m_\rho$ with the IR cut.
The data at the far left side, i.e., $\Lambda_{\rm UV}/a_p=0$, is the standard lattice result.
As in the case of the UV cut, the pion mass and the $\rho$-meson mass decrease by removing the IR gluon.
In addition, in contrast to the case of the UV cut, these meson masses degenerate in $\Lambda_{\rm IR}/a_p\ge 2$.
In the calculation of the interquark potential, the confinement potential vanishes in this energy region of $\Lambda_{\rm IR}$.
Therefore, this degeneracy in meson masses suggests that the quark and antiquark in such mesons become unbound or, if possible, very narrowly bound.
This state is called as ``quasi-free", which implies that, in this state, the Coulomb interaction exists, but the confinement potential does not exist.

\subsection{Staggered fermion}
The staggered fermion is another well-known fermion in lattice QCD, which has a kind of chiral symmetry.
This fermion is often used for investigating the chiral property of hadrons \cite{Ko75,Ba85}.

The staggered fermion calculation is performed on $16^3\times 32$ lattice with $\beta=6.0$.
Then, in the spatial direction, the momentum-space lattice spacing is $a_p \simeq 0.77$ GeV, and in the temporal direction, the momentum-space lattice spacing is $a_p/2$.
The bare current quark mass $m$ in the staggered fermion action is taken as $ma=0.01$, 0.02, 0.06. 0.10, and the corresponding pion masses are about 0.48 GeV, 0.67 GeV, 1.2 GeV, and 1.5 GeV, respectively.

\begin{figure}[t]
\begin{center}
\includegraphics[scale=1.15]{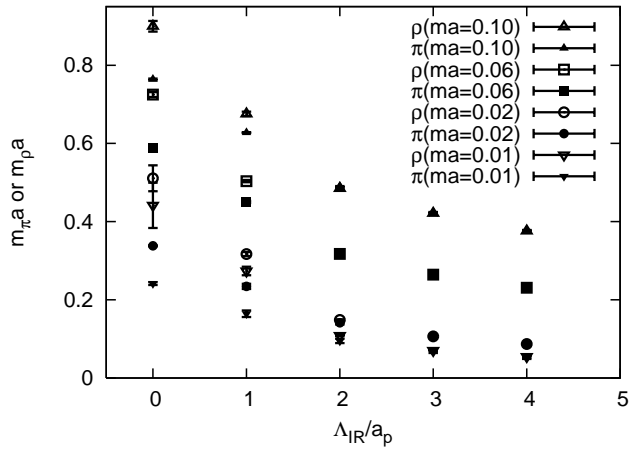}
\caption{\label{Fig9}
The $\Lambda_{\rm IR}$-dependence of meson masses with the staggered fermion.
The bare current quark mass is taken as $ma=0.01$, 0.02, 0.06, and 0.10.
The unit is scaled with the lattice spacing, $a\simeq 0.10$ fm or $a_p \simeq 0.77$ GeV.
}
\end{center}
\end{figure}

The dependence of the meson masses on the IR cut is shown in Fig.~\ref{Fig9}.
The staggered fermion exhibits the same behavior as the clover fermion, i.e., both $m_\pi$ and $m_\rho$ decrease by the IR cut, and they degenerate in $\Lambda_{\rm IR}/a_p\ge 2$.

Next, we consider the dependence of the meson masses on the quark mass.
For chiral extrapolation, the squared meson masses are fitted with quadratic functions as
\begin{eqnarray}
 (m_\pi a)^2 \ {\rm or} \ (m_\rho a)^2 = c_2 (ma)^2 + c_1(ma) +c_0,
\end{eqnarray}
where $c_2$, $c_1$, and $c_0$ are fitting parameters. 
In the standard lattice calculation, the masses of $\rho$-meson and many other mesons are linear functions of the quark mass, and they are finite value even in the chiral limit $m \to 0$.
On the other hand, the pion exhibits characteristic properties.
It obeys the well-known relation near the chiral limit,
\begin{eqnarray}
f_\pi^2 m_\pi ^2 = - m \langle \bar{q}q \rangle ,
\end{eqnarray}
which is called the Gell-Mann-Oaks-Renner relation.
Here, $\langle \bar{q}q \rangle (= \langle \bar{u}u+\bar{d}d \rangle)$ is the two-flavor quark condensate.
In the chiral limit, the pion becomes massless because it is the Nambu-Goldstone boson associated with chiral symmetry breaking.

\begin{figure}[t]
\begin{center}
\includegraphics[scale=1.15]{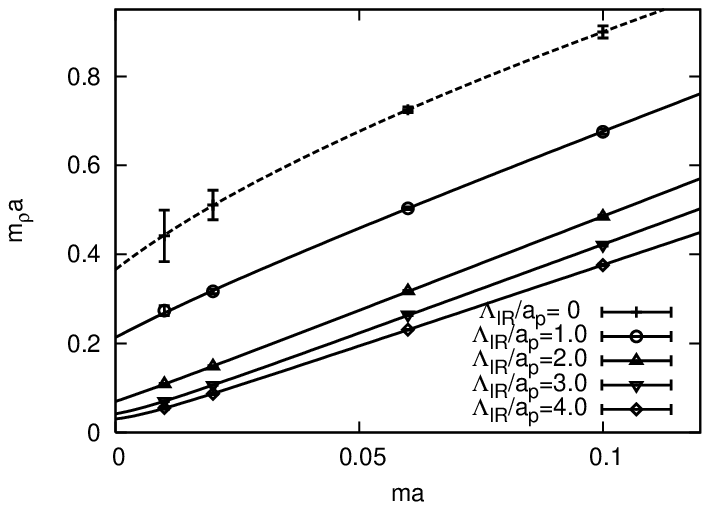}
\caption{\label{Fig10}
The rho mass $m_\rho$ plotted against the bare current quark mass $m$ in the staggered fermion action.
Chiral extrapolation is performed with $(m_\rho a)^2 = c_2 (ma)^2 + c_1(ma) +c_0$.
The data with $\Lambda_{\rm IR}=0$ (broken line) is the standard lattice result.
The unit is scaled with the lattice spacing, $a\simeq 0.10$ fm or $a_p \simeq 0.77$ GeV.
}
\includegraphics[scale=1.15]{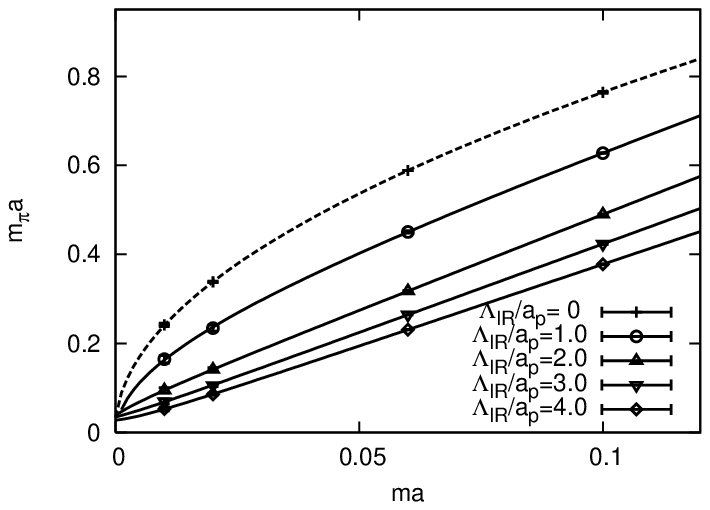}
\caption{\label{Fig11}
The pion mass $m_\pi$ plotted against the bare current quark mass $m$ in the staggered fermion action.
The notation is the same as Fig.~\ref{Fig10}.
}
\end{center}
\end{figure}

We show the quark mass dependence of the pion mass in Fig.~\ref{Fig10}, and that of the $\rho$-meson mass in Fig.~\ref{Fig11}.
When the IR cut is introduced, the $\rho$-meson mass uniformly decreases with the linear extrapolation form unchanged.
In contrast, the pion loses its character in $\Lambda_{\rm IR}/a_p\ge 2$, where the pion mass degenerates to the $\rho$-meson mass.
The extrapolation function becomes a linear function.
In other words, the Gell-Mann-Oaks-Renner relation is broken, and the pion is no longer the Nambu-Goldstone boson.
We note that this behavior of the pion originates from the vanishing of confinement, rather than chiral symmetry restoration.
In $\Lambda_{\rm IR}/a_p\ge 2$, the pion is not a bound state and far from the Nambu-Goldstone boson.

Chiral symmetry breaking is another important topic in nonperturbative QCD. 
It is interesting to accurately investigate chiral properties with a small quark mass and unquenched calculation.

\section{Analytical model calculation}
\subsection{Richardson potential}
From the lattice QCD results, it is found that the relevant energy scale of color confinement is below 1.5 GeV.
In this section, we introduce the IR cut to the phenomenological interquark potential which includes the confinement potential, and compare it with the lattice result.
For this purpose, we analyze the Richardson potential \cite{Ri79}.

The Richardson potential is the phenomenological interquark potential constructed so as to reproduce the Coulomb plus linear structure.
This potential is defined by the one-dressed-gluon-exchange amplitude which is proportional to 
\begin{eqnarray}
\tilde{V}(p^2)=-C_F\frac{g^2(p^2)}{p^2},
\end{eqnarray}
where
\begin{eqnarray}
g^2(p^2)=\frac{1}{\beta_0 \ln(1+p^2/\Lambda^2)}.
\end{eqnarray}
Here, $C_F=4/3$ and $\beta_0=(11N_c-2N_f)/48\pi^2$.
$\Lambda$ is the only parameter in this model.
This coupling is similar to the standard QCD coupling, except for 1 in the argument of logarithm.
After the integration of the time component, the Richardson potential is obtained by the three-dimensional Fourier transformation as
\begin{eqnarray}
V(R)&=&\int \frac{d^3p}{(2\pi)^3}\ e^{i\vec{R}\cdot \vec{p}}\ \tilde{V}(\vec{p}^2)\nonumber\\
&=&\frac{C_{\rm F}}{8\pi \beta_0}\left[ \Lambda^2 R - \frac{1}{R} + \frac{f(\Lambda R)}{R} \right],
\label{VRi}
\end{eqnarray}
where
\begin{eqnarray}
f(x)=4\int_{1}^{\infty}dt \frac{e^{-tx}}{t}\frac{1}{[\ln(t^2-1)]^2+\pi^2}.
\end{eqnarray}
We set $N_f=0$ to compare this potential with the $Q\bar Q$ potential in quenched lattice QCD, and $\Lambda=0.48$ GeV so that the coefficient of the linear potential is equal to the physical string tension $\sigma \simeq 0.89$ GeV/fm.

\subsection{Richardson potential with IR cut}
We introduce the IR cut $\Lambda_{\rm IR}$ to the Richardson potential.
We consider following two ways for the IR cut.

One is the simple IR cut by the three-momentum length on the Fourier transformation,
\begin{eqnarray}
V(R)=\int_{|\vec{p}|\ge \Lambda_{\rm IR}} \frac{d^3p}{(2\pi)^3} \ e^{i\vec{R}\cdot \vec{p}}\ \tilde{V}(\vec{p}^2).
\end{eqnarray}
The other is the change of the functional form as
\begin{eqnarray}
\tilde{V}(\vec{p}^2)=-C_F\frac{g^2(\vec{p}^2)}{\vec{p}^2+\Lambda_{\rm IR}^2}.
\label{VRiIR1}
\end{eqnarray}
The advantage of the latter way is that we can analytically calculate the momentum integral.
The analytical derivation is given in Appendix.
The result is
\begin{eqnarray}
V(R)=\frac{C_{\rm F}}{8\pi \beta_0} \left[ -\frac{2}{R} \Big(\frac{1}{\lambda^2}+h_\lambda e^{-\lambda \Lambda R}\Big) + \frac{f_\lambda (\Lambda R)}{R} \right], 
\label{VRiIR2}
\end{eqnarray}
where
\begin{eqnarray}
h_\lambda &=& \Bigg\{
\begin{array}{cc}
\frac{\ln (\lambda^2-1)}{[\ln (\lambda^2-1)]^2+\pi^2} & (\lambda >1)\\
\frac{1}{\ln (1- \lambda^2)} & (1 \ge\lambda\ge 0)
\label{hl}
\end{array}\\
f_\lambda(x)&=&4P\int_{1}^{\infty}dt \ \frac{te^{-tx}}{t^2-\lambda^2} \frac{1}{[\ln(t^2-1)]^2+\pi^2}
\label{fl}
\end{eqnarray}
and $\lambda=\Lambda_{\rm IR}/\Lambda$.
The symbol $P$ indicates the principal value of the integral, which is necessary in the case of $\lambda>1$.
This functional form becomes the original Richardson potential in the limit of $\Lambda_{\rm IR}\to 0$, apart from an irrelevant constant.

\begin{figure}[t]
\begin{center}
\includegraphics[scale=0.95]{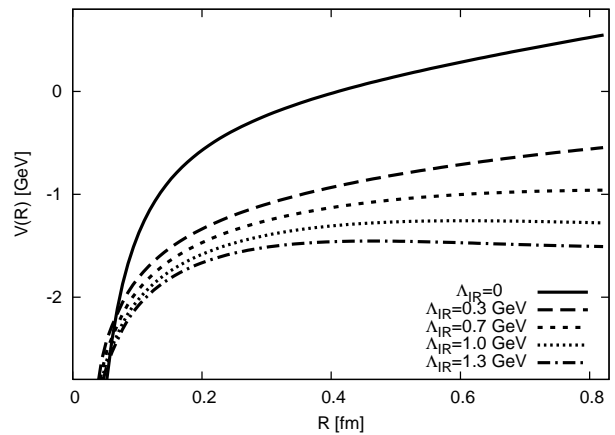}
\caption{\label{Fig12}
The Richardson potential with the IR cut $\Lambda_{\rm IR}$.
The solid line is the original Richardson potential (\ref{VRi}).
The irrelevant constant is arbitrarily subtracted.
}
\end{center}
\end{figure}

These two ways to introduce the IR cut yield almost the same results.
The result of the former way is shown in Fig.~\ref{Fig12}.
The irrelevant constant is arbitrarily subtracted in the figure.
As in the case of the lattice result, the string tension decreases when the IR cut is introduced.
In $\Lambda_{\rm IR}\ge 1$ GeV, the linear confinement potential disappears and the interquark potential becomes a short-range Coulomb-like potential.
This behavior is consistent with our lattice result.
While the Richardson potential is only a phenomenological model and its confinement potential is set by hand, it well reproduces the $\Lambda_{\rm IR}$-dependence of the interquark potential obtained by lattice QCD.

\section{Summary and Discussion}
We have formulated the lattice framework to study the relevant energy scale of QCD phenomena.
We cut link variables in a certain region in momentum space, and calculate the expectation value of physical quantities with the link variables with the cut.
Using this framework, we can determine the gluonic relevant energy scale of QCD quantities.
This framework is broadly applicable for all lattice QCD calculations.

From the asymptotic string tension of the $Q\bar Q$ potential, we have found that the relevant gluonic energy scale of color confinement is below 1.5 GeV.
In fact, the string tension is almost unchanged even after cutting off the high-momentum gluon component above 1.5 GeV.
When we cut the low-momentum component below 1.2 GeV, the confinement potential asymptotically disappears.
Then, when we cut the low-momentum component up to 1.5 GeV, almost all the contributions to the string tension disappear in the whole range of the interquark distance.
In addition, by cutting the infrared gluon in this energy region, the two quarks in mesons are not confined and become quasi-free.

It is often believed that the typical energy scale of nonperturbative QCD is $\Lambda_{\rm QCD}$.
However, the value of the QCD running coupling constant is large even above $\Lambda_{\rm QCD}$.
In fact, our result suggests that the relevant energy scale of color confinement is larger than $\Lambda_{\rm QCD}$.
This fact would be useful as a reference for developing effective theories.
For example, the relevant energy scale determines the cutoff in a low-energy effective model.
Although the value of the cutoff should be based on some physical reasoning, its microscopic derivation is difficult in many cases.
Since lattice QCD is the first-principle calculation in QCD, the relevant energy scale obtained here would provides a strong physical reasoning for the cutoff value.
If we set the ultraviolet cutoff to be 1.5 GeV in momentum integral, we can safely pick up the contribution to the confinement.
Further, the relevant energy scale determines degrees of freedom which appear in the effective field theory \cite{Br05}.

The relevant energy scales can differ even among many nonperturbative phenomena.
Our framework is considered to be a useful tool for understanding the energy scales of many QCD phenomena.

\section*{Acknowledgements}
A.~Y.~and H.~S.~are supported by a Grant-in-Aid for Scientific Research [(C) No.~20$\cdot$363 and (C) No.~19540287] in Japan.
This work is supported by the Global COE Program, ``The Next Generation of Physics, Spun from Universality and Emergence".
The lattice QCD calculations are done on NEC SX-8R at Osaka University.

\appendix
\section{Analytical derivation}

\begin{figure}[b]
\begin{center}
\includegraphics[scale=0.6]{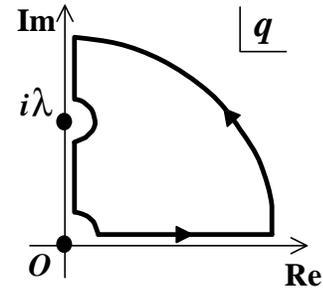}
\caption{\label{Fig13}
Contour in the complex $q$-plane.
}
\end{center}
\end{figure}

In this section, we derive Eq.~(\ref{VRiIR2}) from the Fourier transformation of Eq.~(\ref{VRiIR1}),
\begin{eqnarray}
V(R)=-\frac{C_F}{\beta_0}\int \frac{d^3p}{(2\pi)^3}\frac{e^{i\vec{R}\cdot \vec{p}}}{(\vec{p}^2+\Lambda_{\rm IR}^2)\ln (1+\vec{p}^2/\Lambda^2)}.
\end{eqnarray}
After performing the angular integration, we obtain
\begin{eqnarray}
V(R)=-\frac{C_F}{2\pi^2 \beta_0} \frac{\Lambda}{r}\ {\rm Im} \int_0^{\infty} dq \frac{qe^{irq}}{(q^2+\lambda^2)\ln (1+q^2)},
\label{A2}
\end{eqnarray}
where
\begin{eqnarray}
r\equiv\Lambda R,  \quad  q\equiv |\vec{p}|/\Lambda,  \quad  \lambda\equiv\Lambda_{\rm IR}/\Lambda.
\end{eqnarray}
For performing the integration in Eq.~(\ref{A2}), we consider a contour integral in the complex $q$-plane shown in Fig.~\ref{Fig13}.
The radial part drops with an infinitely large radius.
The two pole contributions at $q=0$ and $i\lambda$ are given by
\begin{eqnarray}
I_0&=& \frac{i\pi}{2}\frac{1}{\lambda^2}\\
I_{i\lambda}&=&  \Bigg\{
\begin{array}{cc}
\frac{i\pi}{2}\frac{e^{-\lambda r}}{\ln (\lambda^2-1)+i\pi} & (\lambda >1)\\
\frac{i\pi}{2}\frac{e^{-\lambda r}}{\ln (1- \lambda^2)} & (1 \ge\lambda\ge 0),
\end{array}
\end{eqnarray}
respectively.
The integral along the imaginary axis is given as
\begin{eqnarray}
&&P\int_{0}^{i\infty}dq \ \frac{qe^{irq}}{(q^2+\lambda^2)\ln (1+q^2)}\nonumber\\
=&&-P\int_{1}^{\infty}dt \ \frac{te^{-rt}}{(\lambda^2-t^2)[\ln (t^2-1)+i\pi]}\nonumber\\
&&-P\int_{0}^{1}dt \ \frac{te^{-rt}}{(\lambda^2-t^2)\ln (1-t^2)}.
\end{eqnarray}
The symbol $P$ indicates the principal value of the integral.
By inserting these, we get
\begin{eqnarray}
V(R)=\frac{C_{\rm F}}{8\pi \beta_0}\frac{\Lambda}{r} \Big[ - \frac{2}{\lambda^2}-2h_\lambda e^{-\lambda r}+ f_\lambda (r) \Big],
\end{eqnarray}
where $h_\lambda$ and $f_\lambda (r)$ are given as Eq.~(\ref{hl}) and Eq.~(\ref{fl}), respectively.

\end{document}